\documentstyle[preprint,aps]{revtex}
\begin{document}
\draft

\title{Magnetic Field Dependence of the Level Spacing of a Small
Electron Droplet} \author{O. Klein \footnote{Present address: Laboratoire
des Solides Irradi{\'e}s, Ecole Polytechnique, F-91128 Palaiseau},
D. Goldhaber-Gordon, C. de C. Chamon, and M.A.~Kastner}
\address{Department of Physics, Massachusetts Institute of
Technology,\\ 77 Massachusetts Ave., Cambridge MA 02139}
\date{Submitted to Phys. Rev. B Rapid Comm.} \maketitle
\begin{abstract}

The temperature dependence of conductance resonances is used to
measure the evolution with the magnetic field of the average level
spacing $\Delta\epsilon$ of a droplet containing $\sim 30$ electrons
created by lateral confinement of a two-dimensional electron gas in
GaAs. $\Delta\epsilon$ becomes very small ($< 30\mu$eV) near two
critical magnetic fields at which the symmetry of the droplet changes
and these decreases of $\Delta\epsilon$ are predicted by Hartree-Fock
(HF) for charge excitations. Between the two critical fields, however,
the largest measured $\Delta\epsilon= 100\mu$eV is an order of
magnitude smaller than predicted by HF but comparable to the Zeeman
splitting at this field, which suggests that the spin degrees of
freedom are important.

\end{abstract}
\pacs{PACS numbers: 73.20.Dx, 73.20.Mf}

\narrowtext

In a recent Letter, Klein {\it et al.} \cite{klein} showed that a
droplet of electrons created by lateral confinement of a
two-dimensional electron gas undergoes changes in its symmetry at two
critical magnetic fields. On one hand, the exchange and the
confinement energies favor a compact electron distribution, on the
other hand, the Coulomb repulsion favors a diffuse occupation. The
magnetic field $B$ alters the balance between these two effects,
causing the ground state (GS) to change \cite{chamon}. Correlations
are also expected to play an important role \cite{jain:3}. Such
symmetry breaking is expected to be accompanied by low energy
excitations. In particular, Hartree-Fock (HF) \cite{chamon}
calculations, which predict these changes in symmetry, also predict
that these are accompanied by a decrease of the average excitation
energy of the droplet. We report, here, measurements of the average
level spacing $\Delta\epsilon(B)$ which confirm these qualitative
predictions. Between the two critical magnetic fields, however, the
largest measured $\Delta\epsilon$ is much smaller than predicted by
current theories for charge excitations. The maximum
$\Delta\epsilon(B)$ is close to the Zeeman energy, which suggests that
spin excitations are important.

The device we use consists of a heavily $n$-doped GaAs substrate
covered by a first layer of AlGaAs and a top layer of GaAs
\cite{meirav}, both layers grown by molecular-beam epitaxy. The strong
electric field created by the the band offset between AlGaAs and GaAs,
as well as the positive voltage $V_b$ applied to the substrate on the
bottom, creates a two-dimensional electron gas (2DEG) at the
AlGaAs/GaAs interface. Ti-Au electrodes are deposited on the top
surface 100nm above the 2DEG and fashioned by electron beam
lithography. A negative voltage maintained constant during the
experiment, is applied to the top electrodes. This confines the
electrons laterally in a potential that is approximately a
two-dimensional harmonic oscillator characterised by the energy
$\hbar\omega_0$ \cite{meirav}. In addition, the electrodes define two
potential barriers, through which electrons must tunnel to enter and
exit the droplet. The 2DEG regions outside the left and right barriers
form the leads. The current through the droplet is measured as a
function of the voltage $V_b$, as well as that between the left and
right leads $V_{lr}$.

Figure 1(a) illustrates the conductance $G$ through the droplet for
$V_{lr}<k_BT/e$, where $G$ is independent of $V_{lr}$. Sharp peaks
arise from resonant tunneling: the $N$th peak occurs at a voltage
$V_N$ such that $e V_N$ is proportional to $(E_{N}-E_{N-1})$, the
energy difference between a ($N$)-particle and ($N-1$)-particle GS. At
high temperatures, this difference is dominated by the Coulomb
charging energy of the droplet. At very low temperatures, however,
there is a quantum mechanical contribution to $(E_{N}-E_{N-1})$
arising from the confinement of the droplet to a small region of
space. When the GS of the droplet changes, $(E_{N}-E_{N-1})$ changes
as well and thus $V_N$ provides a spectroscopic probe of the GS energy
of the droplet.

Figure 1(b) illustrates how this addition spectroscopy
\cite{klein,mceuen:prb,ashoori:prl2} is done by following the
$B$-dependence of the peak position at the base temperature of our
dilution refrigerator corresponding to an electron temperature of 50mK
(see caption of Fig.2). At high $B$, the kinetic energy of the
electrons is quantized in Landau levels (LLs) of index
$n=0,1,...$. The large structure in $V_N(B)$ below 1.5T corresponds to
the transfer of electrons from higher to lower $n$ LLs. The change in
behavior at 1.5T indicates a different phenomenon. Above this field
all the electrons are in the lowest LL (i.e. all electrons have the
same kinetic energy) and the jumps in $V_N$ are caused by the flips of
electron spins. The spins flip, not because of their magnetic moment,
which is small in GaAs ($g=-0.4$), but because of the Coulomb
interaction \cite{mceuen:prb,klein}. The latter scales as
$e^2/\epsilon \ell_B $ where $\ell_B = \sqrt{\hbar c/e B}$ is the
magnetic length and $\epsilon$ is the dielectric constant. Increasing
$B$ decreases $\ell_B$, which increases the Coulomb repulsion, causing
the electrons to spread out \cite{mceuen:prb,klein}.
 
Klein {\it et al.}  \cite{klein} have shown that just below $B_c$
(indicated in Fig.1(b)) the GS of the droplet is a singlet. Their
experiment shows that for a droplet with a small number of electrons
$N\sim 30$ and a large $\hbar \omega_0 \sim 2meV$, the GS has a
compact charge distribution at low fields. The compact occupation of
the lowest LL is achieved when the charge density corresponds to
filling fraction (the ratio of electron to flux-quantum density) equal
to 2 throughout the droplet, {\it i.e.} when both spin states are
equally occupied. Above $B_c$, the growth in the Coulomb repulsion
causes the charge to spread out, causing the electrons to flip their
spins in order to minimize the loss of exchange energy. Thus the
transition at $B_c$ results from a change in symmetry from zero to
finite total spin.

A different kind of symmetry breaking occurs at high fields. Above
$1.9 B_c$ all spins are polarized and the droplet must find a new way
to spread out as $B$ increases. In HF, the GS between $1.9 B_c$ and
$B_r=2.5 B_c$ (Fig.1(c)) is the so-called maximum density droplet
(MDD) \cite{mcdonald,chamon}, the compact charge distribution for a
spin polarized droplet with filling fraction 1 throughout the
droplet. In HF, the symmetry change at $B_r$ is analogous to a
liquid-gas phase transition and the order parameter is the charge
density. Klein {\it et al.}  observe experimentally that $B_r$ is
lower than predicted by HF, as seen by comparing Fig.1(b) and (c),
suggesting that correlations play an important role in this higher
field symmetry breaking \cite{ahn,jain:3}.

HF predicts a dramatic decrease in the average excitation energy
$\Delta\epsilon$ of the droplet near $B_c$ and $B_r$. One way we
determine $\Delta\epsilon$ is from the cross-over from single to
multiple level transport as $T$ is increased \cite{foxman:2}. We
carefully select conductance peaks that have an exponential tail at
base temperature (Fig.1(a)). This indicates that the peak shape is
dominated by the thermal broadening of the energy distribution of the
electrons in the reservoir \cite{foxman:2} and that quantum
fluctuation effects \cite{pasquier} are comparatively small. In
contrast, a Lorentzian tail is the characteristic signature of the
regime where the latter effects are important. Also, we measure the
temperature dependence at $B$ fields well separated from the steps in
Fig.1(b), for at the cusps associated with these we expect degenerate
GS's.

When $k_B T < \Delta\epsilon$, the current is limited by a single
quantum level and the conductance peak profile is given by the
derivative of the Fermi-Dirac distribution function. The data in
Fig.1(a) are well fit by the formula \cite{beenakker}
\begin{eqnarray} G(V_b) = \frac{e^2}{h} \sum_{N=1}^\infty
\frac{\Gamma_N}{4 k_B T} \cosh^{-2} \left( \alpha e \frac{V_b-V_N}{2
k_B T} \right). \label{eq:single} 
\end{eqnarray} 
$\Gamma_N$ is the tunneling matrix element, and the factor $\alpha$
converts a change in $V_b$ to a shift in the electrostatic potential
of the droplet \cite{capa}. Thus, the amplitude of the peak $G_{max}$
decreases as $1/T$ with increasing $T$. However, when $k_B T$ becomes
larger than $\Delta\epsilon$, excited states as well as the GS
participate in the conductance. The number of levels participating
then grows as $T/\Delta\epsilon$, but each channel still contributes a
weight that varies as $1/T$, so the total conductance becomes
temperature independent. The cross-over of $G_{max}$ from $1/T$ to
constant provides a measure of $\Delta\epsilon$. Beenaker
\cite{beenakker} has calculated the line shape in the multi-level
regime assuming a constant $\Delta\epsilon$ and a constant $\Gamma_N$
for all levels in the excitation spectrum of the ($N$)-electron
droplet. He finds that the following is a good approximation:
\begin{eqnarray} G(V_b) \approx \frac{e^2}{h} \sum_{N=1}^\infty 
\frac{\Gamma_N}{2 \Delta\epsilon} \cosh^{-2} \left( \alpha e
\frac{V_b-V_N}{2.5 k_B T} \right). \label{eq:multi} \end{eqnarray} The
cross-over temperature of $G_{max}$ from $1/T$ to constant is at
$\Delta\epsilon /2$.

Concomitant with this cross-over of the peak amplitude, the peak
profile also changes subtly, and this shows up in measurements of the
full width at half maximum (FWHM) \cite{foxman:2}. The FWHM is
proportional to $T$ for both single- and multiple-level transport, but
the FWHM is larger in the multiple-level regime. The increase of the
FWHM also provides an indication of when $k_BT$ is comparable to
$\Delta\epsilon$.

Figure 2(a) shows the $T$-dependence of the conductance peaks at
$B/B_c=1.6$. The inverse of $G_{max}$ as a function $T$ is plotted in
Fig.2(b). The cross-over from $1/T$ to constant is quite clear. The
FWHM as a function of $T$ for one of the peaks in Fig.2(a) is plotted
in Fig.2(c). At $T\approx0.5$K where the amplitude cross-over takes
place, the FWHM also deviates from its low $T$-linear behavior.

Small differences from the model of Eqs.(\ref{eq:single}) and
(\ref{eq:multi}) are observed in Fig.2(b). We observe that $1/G_{max}$
decreases slightly with increasing $T$ above 600mK. This can be
explained by taking into account the variation of $\Gamma$ and
$\Delta\epsilon$ with excitation energy for fixed $N$. One expects
that levels of increasing excitation energy have smaller
$\Delta\epsilon$ and larger $\Gamma$ because of deviations from
parabolicity of the confining potential and because of narrowing of
the two tunnel barriers at higher energy.

Figures 2(d-f) show analogous measurements at $B/B_c=2.1$ close to
$B_r$. Clearly $\Delta\epsilon$ is smaller near the field at which the
symmetry breaking occurs. The $T$-dependence of the FWHM is also
consistent with a small $\Delta\epsilon$. The FWHM is linear in $T$
down to the lowest temperature measured, with a slope larger than the
one measured in Fig.2(c), suggesting that the droplet is already in
the multiple-level regime at the lowest temperatures.

Figure 3(a) shows the $B$-field dependence of $\Delta\epsilon$
extracted from data like those in Fig.2 for devices with 30 and 50
electrons in the droplet. Although the values of $B_c$ differ by 0.2T
between the two cases, the experimental results for the two droplets
are consistent when plotted as a function of $B/B_c$. The level
spacing nearly vanishes at $B/B_c=1$ and 2.2 where the two symmetry
changes occur. Between these two critical fields, $\Delta\epsilon$
reaches a maximum value of 100$\mu$eV.

The inset of Fig.3(b) illustrates the HF prediction for the level
spacing. We have calculated the gap between the GS and the lowest
excited state as a function of $B$. The latter energy vanishes at
level crossings, which occur near steps in Fig.1(c), and rises to a
maximum between steps. We have plotted, in the inset of Fig.3(b), the
lines that join all these local maxima. In the experiment, we have
also measured $\Delta\epsilon$ between steps. The overall shape of
Fig.3(a) is reproduced by HF: $\Delta\epsilon$ vanishes near $B_c$,
rises to a maximum, and then falls again precipitously near
$B_r$. However, the largest values of $\Delta\epsilon$ predicted by HF
are ten times larger than observed experimentally.

Another way to measure $\Delta\epsilon$ is to use tunneling excitation
spectroscopy (TES) \cite{foxman:prb,johnson}. In this scheme, the
differential conductance is measured as a function of $V_{lr}$ and the
quantum levels are seen as peaks. Two spectra are shown in Fig.4. At
$B/B_c=1.6$, where $\Delta\epsilon$ is largest, the level spacing
measured this way agrees well with that measured from the
$T$-dependence. However at $B/B_c=2.1$, close to $B_r$, TES gives a
value somewhat larger than does the temperature dependence, although the
value is much smaller than at $B/B_c=1.6$. TES measures the
single-particle excitations \cite{daniela} while the $T$-dependence
also includes many-body excitations. This may be the reason for this
discrepancy.

The observation of very small $\Delta\epsilon$ near $B_c$ and $B_r$
confirms the interpretation of Klein {\it et al.} \cite{klein} that
symmetry breaking occurs at these critical fields. However the
observation that the largest $\Delta\epsilon$ is much smaller than
the value predicted by HF calls into question the nature of the
low-lying excitations of the MDD.

The existing models that describe the low-lying excitations near the
MDD can be separated in two categories depending on whether they
consider charge or spin excitations. Consider, first, the former,
which assumes that both GS and excited states are spin polarized. This
class includes the HF calculation of Chamon and Wen \cite{chamon},
which uses as a basis the single-particle states of the symmetric
gauge, without level mixing. The single-particle states are labeled by
an angular momentum index $m>0$, and they represent circular orbits of
radius $\ell_B \sqrt{2(m+1)}$. In the MDD GS, all the innermost
orbitals are occupied ($m=0,1,..,N-1$) and the charge density has the
spatial distribution corresponding to filling fraction 1 uniformly
over the droplet. This compact charge distribution is
incompressible. As $B$ increases, however, the HF charge distribution
approaches the classical dome-like shape corresponding to a state
which is compressible throughout the droplet. The first step in this
transformation is the edge reconstruction at $B_r$ where holes are
first introduced in the interior of the droplet. This transition, from
an incompressible state to a compressible one, gives rise to the
abrupt decrease of the level spacing at $B_r$. The single-particle
level spacing in the compressible state can be evaluated from the
expression $2 \hbar^2 /m^\star r^2 g_s \sim 40\mu$eV where $m^\star$
is the effective mass, $g_s=1$ is the spin-degeneracy and $r=200$nm is
the radius of the MDD. A better approximation than HF is the model of
Oaknin {\it et al.} \cite{oaknin} in which charge magnetoexcitons are
the low-lying excitations. These are single electron-hole pair
excitations of the MDD which correspond to moving an electron from the
interior to the exterior of the droplet. While such excitons give rise
to excited states in the MDD, they become stable in the GS above
$B_r$. Above $B_r$, as in HF the electron occupancy is reduced within
a few $\ell_B$ inside the edge of the MDD. Other models that
incorporate correlations include the work of Kamilla and Jain
\cite{jain:2,jain:3} that study excitations of non-interacting
composite fermions. In all the models we have discussed, the level
spacing in the MDD state is of order $e^2/\epsilon \ell_B$, which is
approximately $7$meV near 3T, the $B$-field at which the MDD is formed
for the data in Fig.3(a), and is therefore much larger than the
$\Delta\epsilon$ we observe.

In contrast, the Zeeman energy is 75$\mu$eV at 3T (using $g=-0.4$), a
value closer to our measured $\Delta\epsilon$. Several calculations
have appeared recently on spin-wave like excitations \cite{sondhi} in
a droplet \cite{oaknin:2}. By canting the spins of the electrons
gradually over the MDD, the droplet reduces the cost in exchange
energy of having two neighbouring electrons with opposite spins. Such
states may also be a better description of the GS near $B_r$. Even if
the MDD is a good description of the GS, excited states that involve
spin excitations may have lower energy than those involving charge
excitations, which cost both confinement and exchange energies. In
particular, a uniform rotation of all spins costs no exchange or
confinement energy at all. The total spin quantum numbers of the
spin-polarized MDD are $S^2=(N/2+1)N/2$ and $S_z=N/2$. The first
excited state has the same $S^2$ as the MDD but $S_z=N/2-1$ and the
energy gap is then given by the Zeeman energy. This excited state can
also be obtained by including vertex corrections in the Coulomb
interactions \cite{hawrylak:prb}. While, this gives $\Delta\epsilon$
of the right order of magnitude, the actual excited states may be more
complex, involving an admixture of both spin and charge excitations
depending on $N$ and the Zeeman energy.

We have plotted in Fig.3(b) the lowest $\Delta\epsilon(B)$ predicted
for either spin excitations or charge excitations. The upper branch of the
trapezoidal shape in Fig.3(b) is the Zeeman energy. The two abrupt
decreases at $B_c$ and $B_r$ are the low-lying charge excitations
predicted by HF near the fields at which symmetry changes.

Our measurements confirm the predictions of HF that $\Delta\epsilon$
vanishes near two critical fields at which changes in the symmetry of
the droplet occur. Between these two critical fields, however, the
spacing is of the order of the Zeeman energy suggesting that the
low-lying excitations are those of the total spin of the droplet
rather than charge excitations included in HF and some other models.



We are grateful to U. Meirav for his valuable help in this work. We
wish to thank P. Hawrylak, S. A. Kivelson, A. H. MacDonald, J. Oaknin,
J. J. Palacios, S. L. Sondhi, and X.-G. Wen for many useful
discussions and communications. This work was supported by NSF Grant
No. ECS 9203427 and by the U.S. Joint Services Electronics Program
under Contract No. DAALL03-93-C-0001. D.G-G. acknowledge a fellowship
from the Fannie John Hertz Foundation.



\begin{thebibliography}{10}
\bibitem{capa}
The coupling between the droplet and the four external electrodes (the bottom
  substrate, top electrodes and left/right leads) is represented by four mutual
  capacitances: $C_b = 175$aF, $C_t = 60$aF, $C_l=44$aF and $C_r=120$aF at
  2.7T. In this notation, $\alpha = C_b/\sum_i C_i$, $U = e^2/\sum_i C_i$, and
  $\beta= (1-C_r/\sum_i C_i)$.
\end{thebibliography}

\begin{figure} \caption{
(a) Linear conductance through the droplet as a function of the
voltage on the bottom gate $V_b$ at $B=2.7$T. (b) Variation with $B$
of the position of the peak near 0.160V. (The $B$-sweep rate is 0.025
G/sec.)  We count 13 steps between $B_c$ and $2 B_c$ which means that
$N=$ 26 or 27. (c) HF calculation of the resonant energy as a function
of $B$ for a droplet containing $N=27$ electrons and
$\hbar\omega_0=2.1$meV. The model has the same number of steps and the
same $B_c$ as the experiment. Note that the height of the steps is
larger in HF (c) than in the experiment (b) suggesting that the
excitation gap is also larger in HF.}
\end{figure}

\begin{figure} \caption{
Conductance vs. $V_b$ for $T$ ranging from 100mK to 600mK in
increments of 100mK, (a) measured at $B/B_c=1.6$ and (d) at
$B/B_c=2.1$. (b) and (e) Inverse of the conductance peak vs. $T$ for
the peaks in (a) and (d) respectively. The cross-over from $1/T$ to
constant determines $\Delta\epsilon/2$. (c) and (f) Full width at half
maximum vs. $T$ for one peak in (a) and (d) respectively. The error
bars are determined by comparing the behavior of other peaks. In (c),
the deviation from the low $T$ straight line gives a similar value of
$\Delta\epsilon/2$ as in (b). The width and amplitude at base
temperature indicate that the 2DEG is at 50mK, whereas the base
temperature of the dilution refrigerator is 25mK, measured by nuclear
orientation thermometry.}
\end{figure}

\begin{figure} 
\caption{(a) $\Delta\epsilon$ obtained using the analysis described in
the text as a function of $B/B_c$. The open circles are for a droplet
containing $N\sim 30$ electrons and the solid circles are for one
containing $N\sim 50$ electrons. (b)inset: HF calculation of the
$B$-dependence of the gap between the GS and the first excited
state. The gap vanishes near each of the steps in Fig.1(c) and rises
to a maximum between. The solid line joins all the local maxima
between steps. The dashed line is the Zeeman energy. Main:
$B$-dependence of the smallest of these two curves shown in the
inset. The result illustrates $\Delta\epsilon(B)$ for both charge or
spin excitations.}
\end{figure}

\begin{figure} 
\caption{ Tunneling excitation spectra, $dI/dV_{lr}$ as a function of
$V_{lr}$ at (a) $B/B_c=1.6$ and (b) $B/B_c$=2.1. $U \sim 0.5$meV is
the Coulomb gap and $\Delta\epsilon$ is the level spacing. A factor
$\beta=0.7$ converts the $x$-axis scale to meV units [12].}
\end{figure}

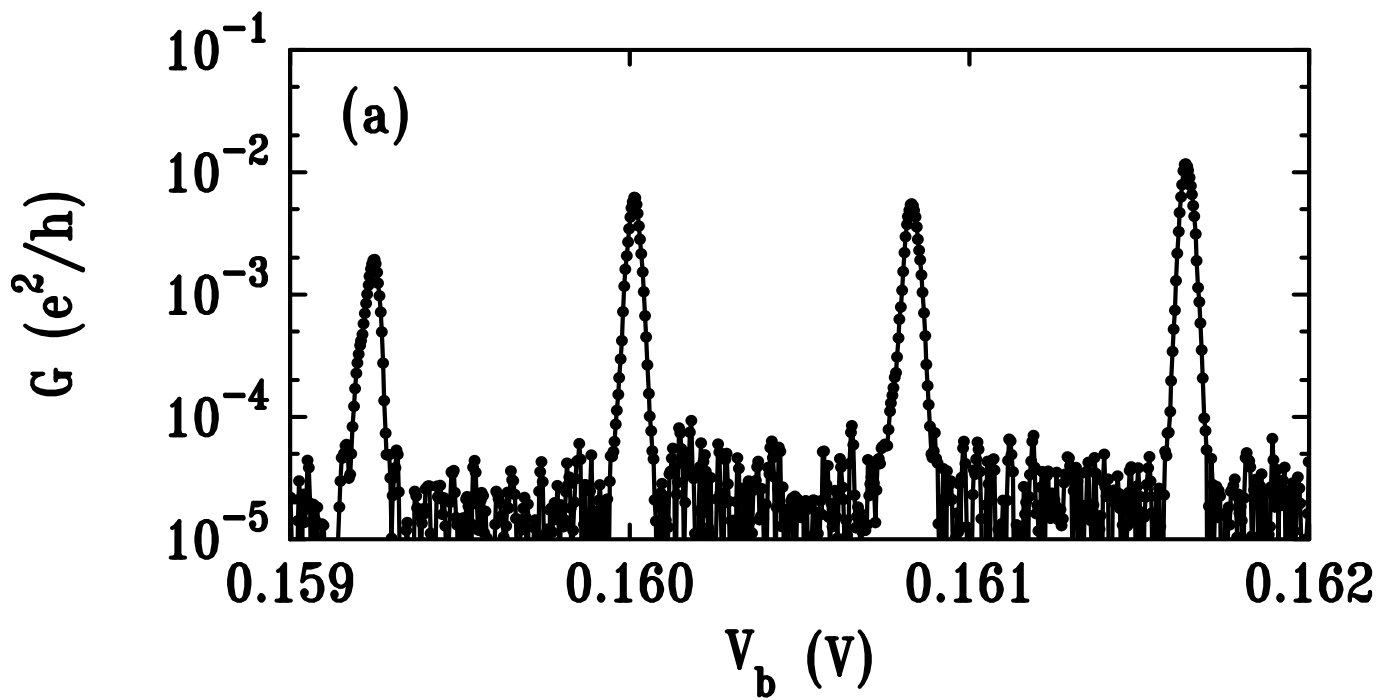
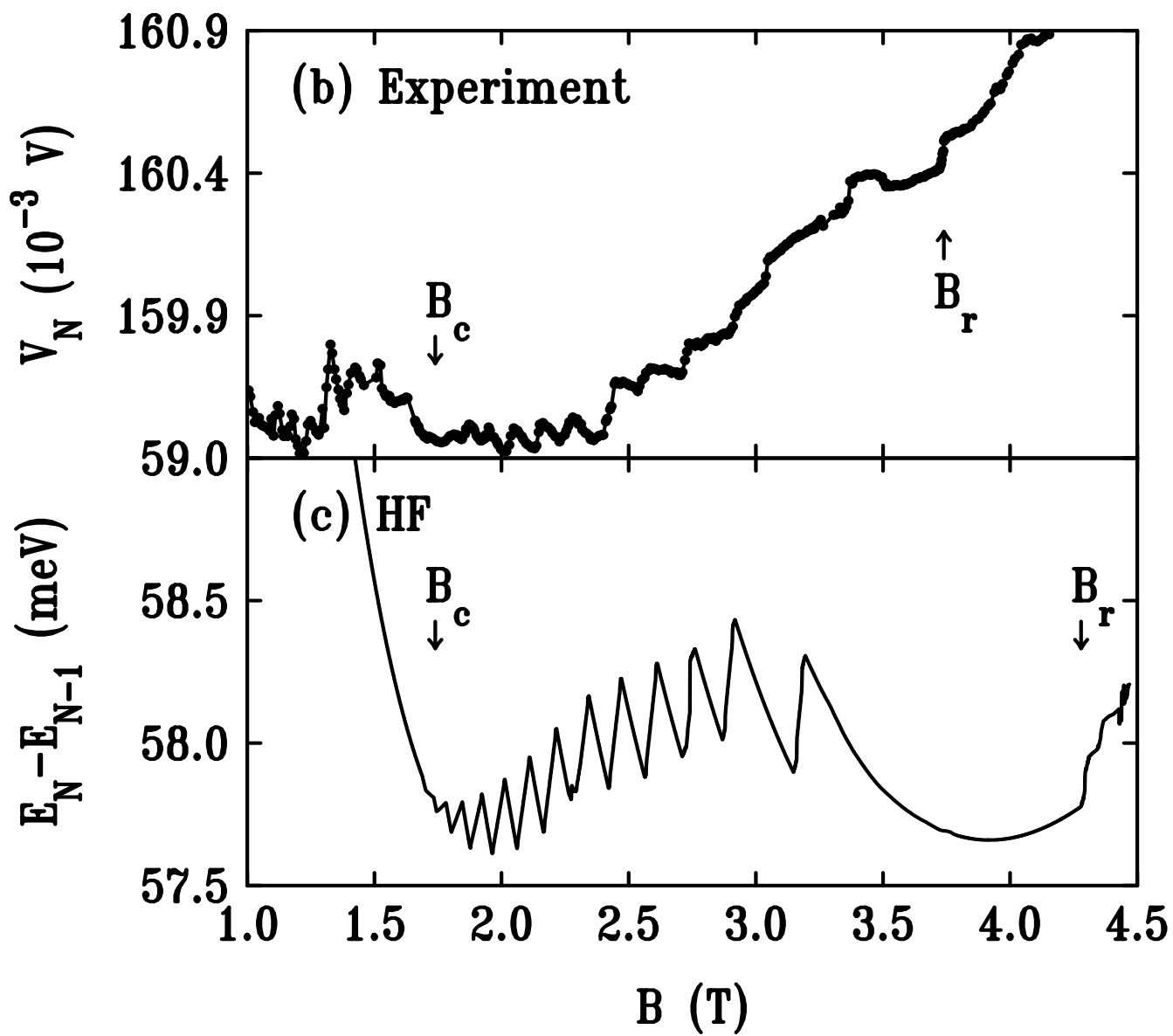
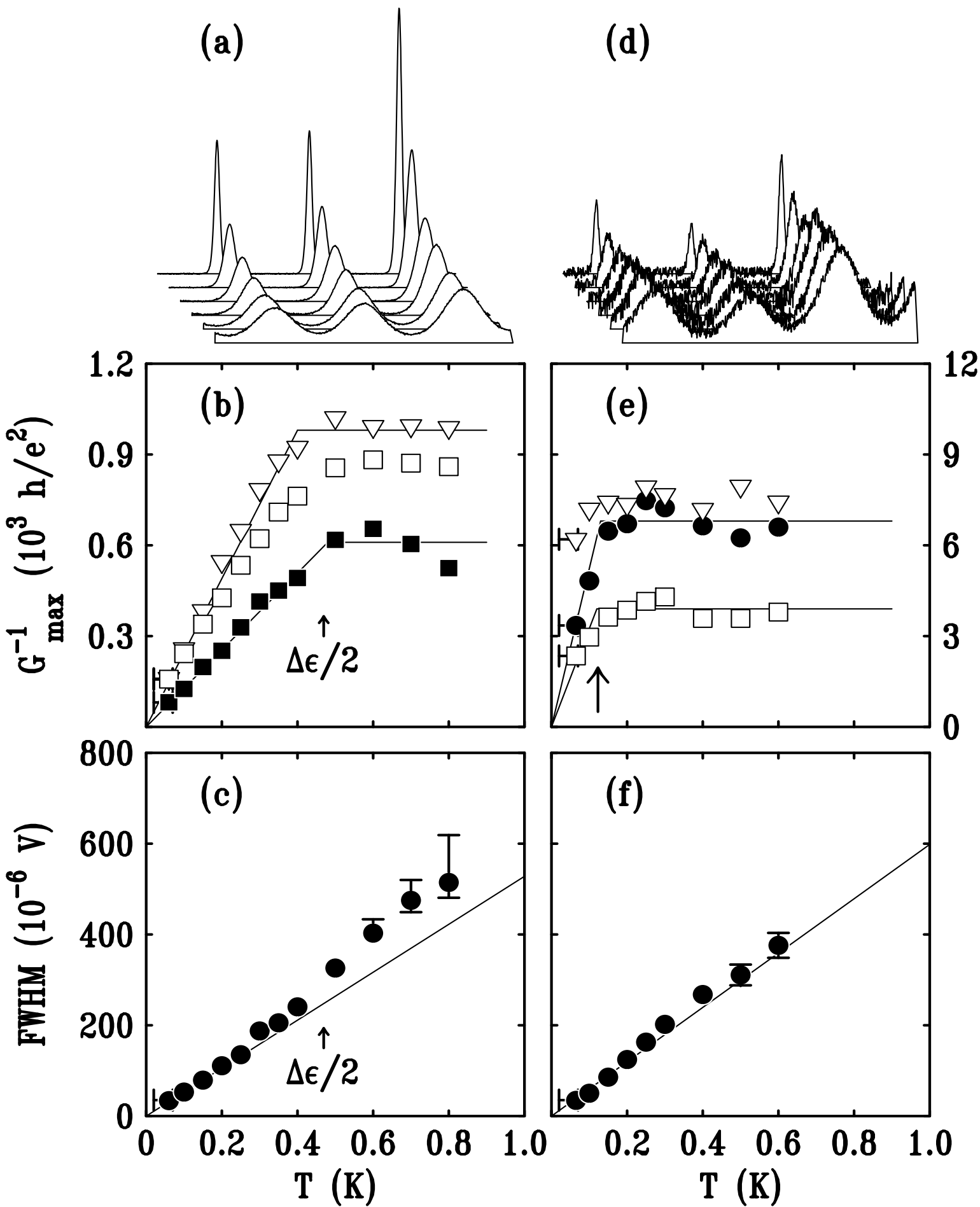
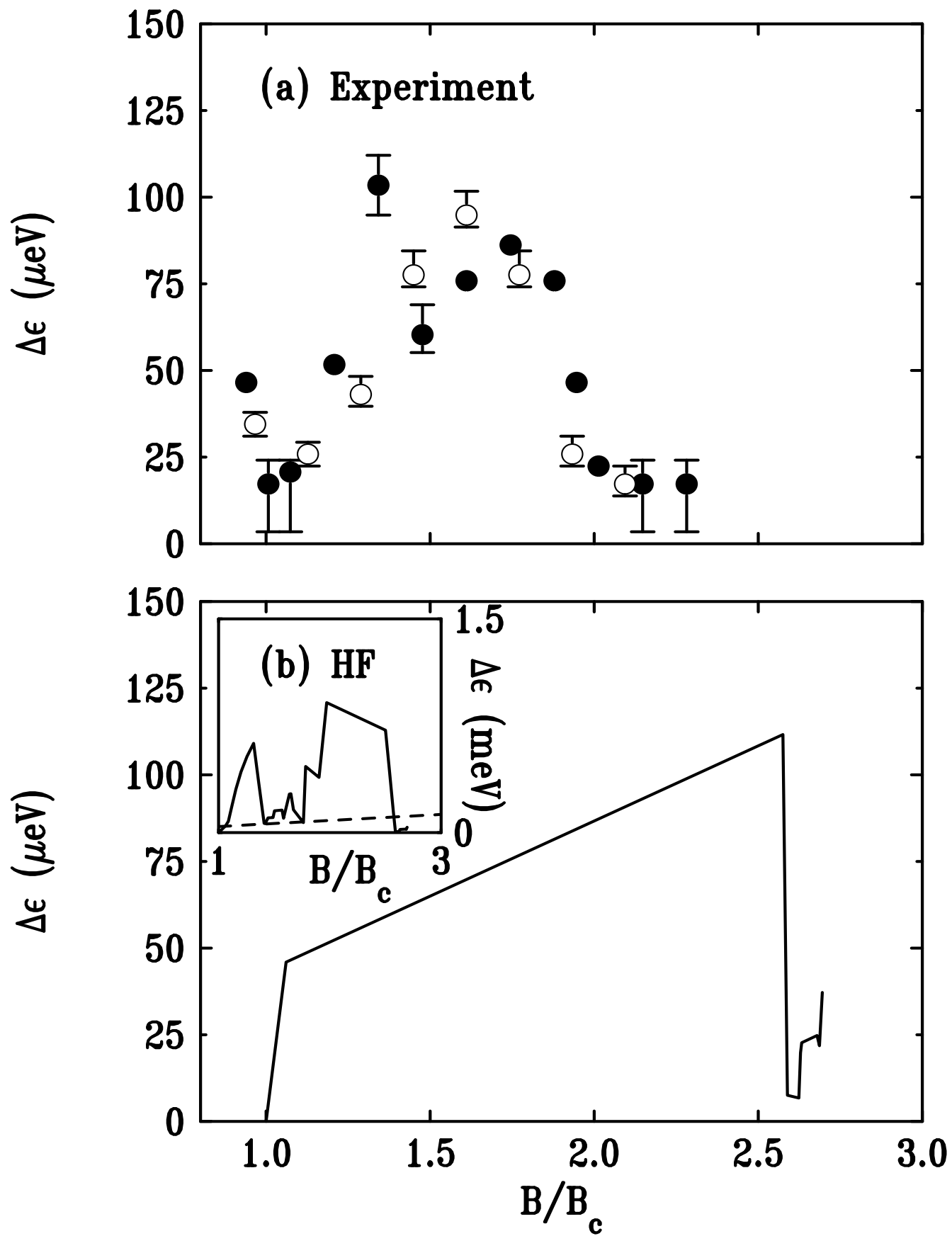
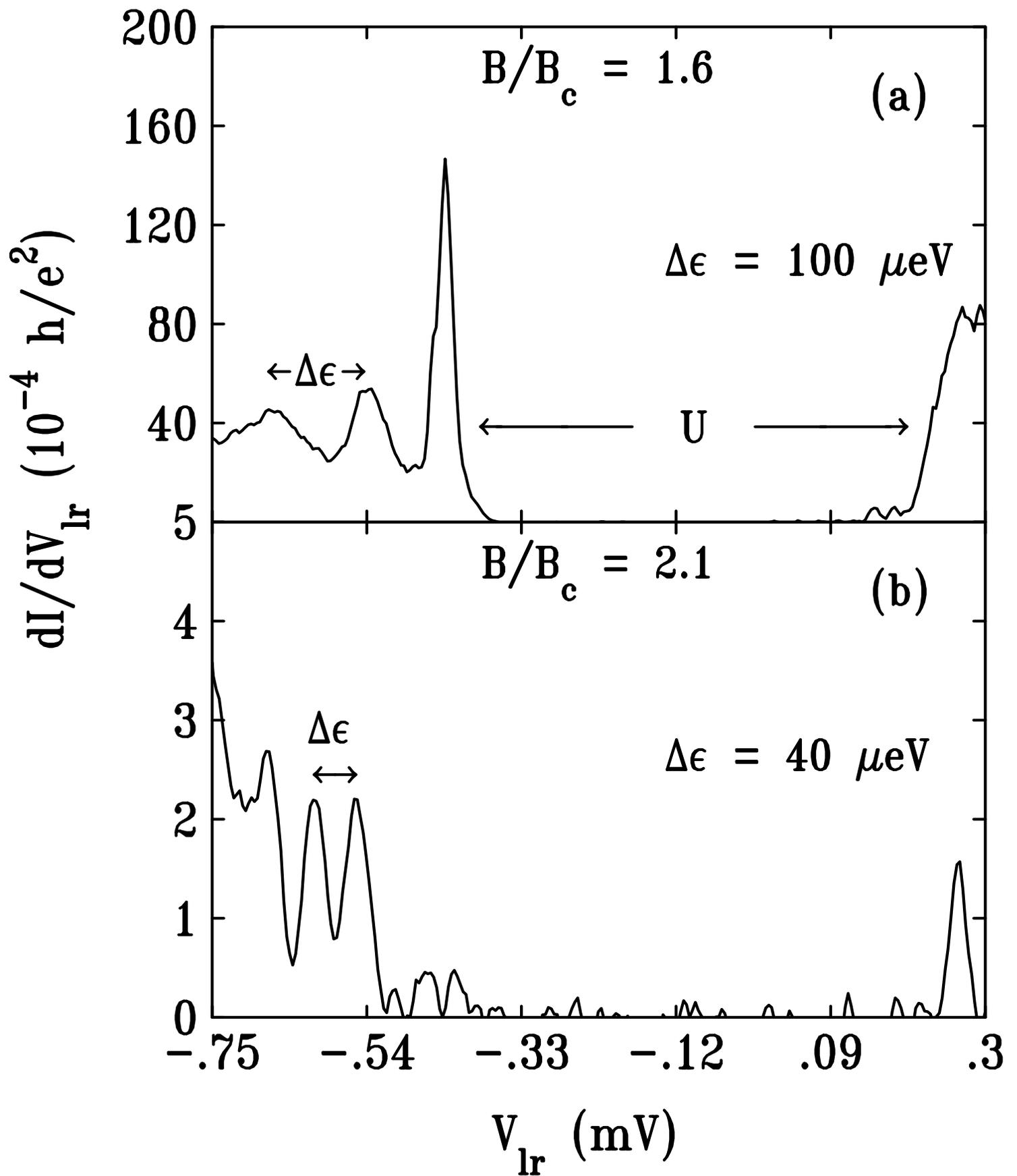
\end{document}